# Preparation, structure and giant magnetoresistance of electrodeposited Fe-Co/Cu multilayers

B.G. Tóth[1]*, L. Péter[1], L. Pogány[1], Á. Révész[2] and I. Bakonyi[1]

[1]*Wigner Research Centre for Physics, Hungarian Academy of Sciences.*
*H-1121 Budapest, Konkoly-Thege út 29-33, Hungary*

[2]*Department of Materials Physics, Eötvös University.*
*H-1117 Budapest, Pázmány Péter sétány 1/A, Hungary*

(Dec. 6, 2013)

**Abstract** – No systematic studies have been carried out on the giant magnetoresistance (GMR) of electrodeposited (ED) Fe-Co/Cu multilayers since the elaboration of a method for the optimization of the Cu layer deposition potential. In this paper, we present results on the electrochemical optimization of the Cu layer deposition potential which was found to depend on the relative iron concentration in the bath. An X-ray diffraction study of ED $Fe_5Co_{95}$(1.5 nm)/Cu($d_{Cu}$) multilayers with $d_{Cu}$ ranging from 0.8 nm to 10 nm revealed an fcc structure. For most of the multilayers, a weak superlattice satellite reflection could be identified. The room-temperature magnetoresistance was studied in detail as a function of the individual layer thicknesses. Multilayers with Cu layer thicknesses above about 1.5 nm were found to exhibit a GMR behavior with a maximum GMR of about 5 % and a typical saturation field of 1 kOe. The GMR magnitude decreased with increasing Fe-content in the magnetic layer. The spacer layer thickness evolution of the MR data was established in detail after separating the ferromagnetic and superparamagnetic GMR contributions and no oscillatory GMR was found. A comparison with literature data on both physically deposited and ED Fe-Co/Cu multilayers is also made.

---

*Corresponding author. E-mail: toth.bence@wigner.mta.hu

**Introduction**

Since bulk Fe-Co alloys have good soft magnetic properties [1], multilayers with this kind of alloy as the magnetic layer can have favorably low coercivity. A detailed giant magnetoresistance study (GMR) study of sputtered TM/Cu multilayers by using binary and ternary alloys of Fe, Co and Ni as magnetic layers has been carried out by Miyazaki et al. [2] who also mapped out the anisotropic magnetoresistance (AMR) of bulk alloys of the Fe-Co-Ni system. As to the GMR of physically deposited Fe-Co/Cu multilayers, it was shown that they can, indeed, have a significant room-temperature GMR [2,3] with fairly low saturation fields [3].

Inomata and Saito [3] investigated sputtered Fe-Co/Cu multilayers with Fe-concentrations of 10, 20 and 50 %. In their study, the Cu layer thickness was varied between 0.9 and 3.8 nm and three oscillatory GMR peaks were found as a function of the spacer layer thickness. Rafaja et al. [4] found about 25 % GMR on sputtered [$Fe_{10}Co_{90}$ (1.1 nm)/Cu(2.2 nm)]x20 multilayers but a detailed study on the influence of the individual layer thicknesses was not carried out. Gangopadhyay et al. [5] also studied sputtered $Fe_{10}Co_{90}$/Cu multilayers. The Cu layer thickness was varied between 0.5 and 3.1 nm and an oscillatory GMR was found, but contrary to the typical result on sputtered multilayers, the second peak was found to be the larger.

There are contradictory experimental data in the literature on physically deposited FM/Cu multilayers, whether a pure Co magnetic layer or a Fe-Co alloy shows larger GMR value. Kataoka et al. [6] and Inomata and Saito [3] present data according to which for [$Fe_{20}Co_{80}$(1.0 nm)/Cu(1.0 nm)]x15 multilayers the GMR is higher than for multilayers with pure Co as magnetic layer, while the work of Miyazaki et al. [7] shows that Co/Cu multilayers exhibit larger GMR than Fe-Co/Cu multilayers.

Whereas there is an abundant literature [8] on the GMR of electrodeposited (ED) Co/Cu, Ni/Cu, Ni-Co/Cu multilayers, reports on the GMR of ED multilayers with Fe in the magnetic layers are much less frequent. Specifically, there have been only two studies on the GMR of ED Fe-Co/Cu multilayers [9,10] and two more papers have only been found for ED Fe-Co/Cu multilayers [11,12]. However, in none of these works the electrochemically optimized Cu-deposition potential was used, which is crucial to eliminate both the codeposition of the magnetic material with Cu and the dissolution of the previously deposited magnetic atoms during the Cu deposition pulse.

The reason for the lack of publications regarding this ED multilayer system is certainly due to the fact that electrochemical deposition of Fe-containing multilayers raises several difficulties. It is hard to electrodeposit pure iron because of its sensitivity to water and oxygen and its high tendency for corrosion compared to nickel and cobalt and because of the instability of electrolytes containing $Fe^{2+}$ ions. This instability originates from the oxidization of $Fe^{2+}$ ions to $Fe^{3+}$ ions and the resulting formation of different precipitates (mostly $Fe(OH)_3$) in the electrolyte. This leads not only to a continuous decrease of the concentration of $Fe^{2+}$ ions available for electrodeposition but also to the appearance of other components in the solution which may then incorporate into the deposit during the deposition. The progress of this process is indicated by the opalescence of the solution, visible even by naked eye. Furthermore, the $Fe^{3+}$ ions in the electrolyte can result in the corrosion of the Cu in the deposit. $Fe^{3+}$ ions present in the solution also deteriorate the current efficiency, in particular when after the $Fe^{3+} + e^- = Fe^{2+}$ reaction, the formed $Fe^{2+}$ ions do not deposit later on in the deposition process. As a result, contrary to electrolytes containing only salts of nickel, cobalt and copper ions, solutions containing iron can only be used for a short time.

Thus, because of the instability of the electrolyte, the electrodeposition of Fe-Co/Cu multilayers was a non-trivial problem to solve, especially at high $Fe^{2+}$ ion concentrations. Furthermore, the optimization of the thickness of the magnetic and the non-magnetic layer to reach the highest possible GMR was also a task to solve. Before doing that, it was necessary to carry out an optimization of the Cu deposition potential [8,13,14] since this can only ensure that the actual layer thicknesses correspond to the nominal ones set during deposition and that the non-magnetic layer will not be contaminated by magnetic atoms. Having the electrochemically optimized Cu deposition potential $E_{Cu}^{EC}$, one can study the dependence of GMR on the true magnetic and non-magnetic layer thicknesses for the first time. This was an important goal in the present study since in the two previous reports on the GMR of ED Fe-Co/Cu multilayers [9,10], such an optimization was not carried out. For a better characterization of our multilayers, X-ray diffraction (XRD) studies have also been performed.

## Experimental details

*Electrolyte for Fe-Co/Cu multilayer deposition.* — For the multilayer deposition, two-pulse plating from a single-bath was applied, i.e., the salts of all metals to be deposited were present in one electrolyte.

To get rid of the effect of precipitates appearing in the solution with time, two stock solutions were prepared. The first one, the Co-solution, was prepared on the basis of the previously elaborated solution designed to deposit pure Co. The second one, the Fe-solution, was prepared with the required composition of the other components but the Fe(II)-sulfate was only added immediately before the electrodeposition experiments. The pH of the Fe-solution was set to 3.25, equal to the pH of the Co-solution. Then, the electrolytes for the deposition of the Fe-Co/Cu multilayers were mixed from these two stock solutions. The concentrations of the two stock solutions were as follows: $CoSO_4 \cdot 7\,H_2O$ or $FeSO_4 \cdot 7\,H_2O$: 0.74 mol/ℓ; $CuSO_4 \cdot 5\,H_2O$: 0.010 mol/ℓ; $Na_2SO_4$: 0.10 mol/ℓ; $H_3BO_3$: 0.25 mol/ℓ; $H_2NSO_3H$: 0.25 mol/ℓ. The mixed electrolytes will be characterized by the relative ion concentration $c_{\text{ion,Fe}}$ in the bath which was defined by the ionic ratio $Fe^{2+}/[Fe^{2+} + Co^{2+}]$.

Iron and cobalt, like the Ni-Co system, show anomalous codeposition [15-18]. However, in this case, Fe is the metal with preferred deposition (nevertheless, in some chloride baths with high chloride ion concentrations, regular codeposition was also found). The anomalous nature of the codeposition process has to be taken into account during the preparation of the electrolytes; i.e., the solutions have to be relatively cobalt-rich.

*Fe-Co/Cu multilayer preparation and characterization.* — The Fe-Co/Cu multilayers were deposited on a [100]-oriented, 0.26 mm thick silicon wafer covered with a 5 nm thick Cr and a 20 nm thick Cu layer, both made by evaporation. The purpose of the chromium layer was to assure adhesion and the Cu layer was used to provide the electrical conductivity of the cathode surface. The deposition was performed in a tubular cell of 8 mm x 20 mm cross section with an upward-facing cathode at the bottom of the cell [13,19]. The deposition was carried out by a galvanostatic-potentiostatic (G/P) pulse combination [8,13]. For the deposition of the magnetic layer, galvanostatic (G) mode was used at -31.25 mA/cm$^2$ current density. For the Cu-layer, potentiostatic (P) mode was used and a saturated calomel electrode (SCE) was used as reference. By varying the deposition time in the G mode, the magnetic layer thickness could be set to a predetermined value. According to our previous experience with Ni-Co alloy electrodeposition from a sulfate bath [20], for these two metals the current efficiency is as high as 96 %. Therefore, for simplicity, a current efficiency of 100 % was assumed for the magnetic layer also in the present ED Fe-Co/Cu multilayers and the preset nominal values were determined from Faraday's law. For controlling the thickness of the Cu layer, the charge flowing through the system was measured during the P pulse. Then, one can calculate the charge necessary to get the preset nominal layer thickness from Faraday's law. The current efficiency for Cu deposition at the optimal potential is usually taken as 100 % since the H$_2$ evolution is negligible; therefore, we used this value also here. Recent detailed X-ray diffraction studies on ED Co/Cu multilayers by using full-profile fitting [21,22] confirmed that both the magnetic and non-magnetic layer thicknesses are within about 10 % of the preset values obtained on the basis of Faraday's law.

Due to the optimization of the Cu deposition potential, the previously deposited Fe-Co alloy layer cannot dissolve during the P pulse. It is ensured this way that both the magnetic and non-magnetic layer will have a thickness as preset from the electrodeposition parameters.

Several sample series were produced with the common goal of investigating the effect of both the Fe/Co ratio in the magnetic layer and the individual layer thicknesses on the electrical transport properties of the samples. For the series with varying Fe/Co ratio in the magnetic layer, the layer thicknesses were $d_{FeCo}$ = 2.5 nm and $d_{Cu}$ = 3 nm with a total multilayer thickness $\Sigma d$ of 100 nm. For the series prepared for studying the layer thickness dependence of the GMR, the magnetic layer thickness was chosen to be 1.0, 1.5, 2.0 and 2.5 nm and the Cu layer thickness was varied between 0.8 and 6.0 nm while $\Sigma d$ was 150 nm.

The overall multilayer composition was measured with electron probe microanalysis (EPMA) in a JEOL JSM-840 scanning electron microscope.

X-ray diffraction (XRD) was used to investigate the structure of some selected Fe-Co/Cu multilayers with the help of a Philips X'pert powder diffractometer in the $\Theta$-$2\Theta$ geometry with Cu-$K_\alpha$ radiation. Using the parameters of the X-ray diffractograms, the lattice parameter and the bilayer repeat length were determined. Since this requires the knowledge of the position of the XRD peaks, the background-corrected and smoothed XRD data were fitted by Lorentzian curves.

*Measurement of electrical transport properties.* — The electrical transport parameters were determined at room temperature by using four-point-in-line probes. The zero-field electrical resistivity ($\rho_0$) of the ED Fe-Co/Cu multilayers was measured in the as-deposited state of the samples, i.e., while still being on their Si/Cr/Cu substrates and before putting them in a magnetic field. The shunting effect of the substrate was decomposed from the measured resistivity values. To obtain the magnetoresistance (*MR*), the resistance was measured as a function of the external magnetic field (*H*) up to 8 kOe. The MR ratio was defined with the formula $MR(H) = [R(H) - R_0]/R_0$ where $R_0$ is the resistance maximum of the sample in a magnetic field close to zero and $R(H)$ is the resistance in an external magnetic field *H*. The magnetoresistance data were determined in the field-in-plane/current-in-plane geometry in both the longitudinal (*LMR*, magnetic field parallel to the current) and the transverse (*TMR*, field perpendicular to the current) configurations. If one takes the difference between the longitudinal and the transverse component, the anisotropic magnetoresistance can be obtained: $AMR = LMR - TMR$.

The measured *MR(H)* curves were decomposed according to a procedure described previously [23] in order to establish the ferromagnetic ($GMR_{FM}$) and superparamagnetic ($GMR_{SPM}$) contributions to the *GMR*. In this process, a Langevin function was fitted to the high-field section of the *MR(H)* curves (beyond the saturation if the FM regions) which yielded the parameters of the $GMR_{SPM}$ contribution and, then, this term was subtracted from the experimental data which procedure provided us the FM contribution to the magnetoresistance. In most cases, a linear term should also be considered in the fitting procedure which accounted for the linearly decreasing resistivity of a ferromagnet with increasing magnetic field at finite temperatures (so-called paraprocess [1]).

## Electrochemical studies for Fe-Co/Cu multilayer deposition

*Electrochemical characterization of the electrolyte.* — For multilayer deposition with reliable layer thicknesses, the electrochemically optimal potential $E_{Cu}^{EC}$ has to be established where neither the dissolution nor the codeposition of either of the magnetic metals can take place. Depending on the experimental parameters, cyclic voltammograms can be used to establish the value of $E_{Cu}^{EC}$ with a limited relevance [14]. The key factor affecting the appearance of an inflection point on the anodic-going scan of the cyclic voltammograms is the coverage of the more noble metal on the magnetic layer by the time when the anodic-going sweep reaches the dissolution potential of the magnetic layer. The onset potential of the dissolution of the magnetic layer can be a function of its composition, too, and the selective dissolution of the least noble element is also possible.

Therefore, in order to obtain some preliminary information for the optimization of the Cu deposition potential, the polarization curves of the solutions with two different $Fe^{2+}$ concentrations were measured, in addition to the pure Co-electrolyte (see Fig. 1). For the pure Co-solution (Fig. 1a), an inflection point can clearly be seen at -0.585 V as observed also in our earlier works [14] because the magnetic layer is a single metal and Co can dissolve nearly reversibly even at slightly more positive potentials than the $E_{Cu}^{EC}$ value.

An inspection of the cyclic voltammogram curves of the two Fe-containing solutions shown in Figs. 1b and 1c suggests that $E_{Cu}^{EC}$ should lie somewhere in the potential range of the extended plateau between approximately -0.8 V and -0.5 V, since the optimal potential where neither the dissolution of the magnetic material, nor the codeposition of the magnetic material with Cu will occur can be expected only in this range.

*Optimization of the Cu deposition potential: current transient study.* — To determine the exact value of $E_{Cu}^{EC}$, the current transients have to be measured over the plateau region [14]. The potential ranges determined from the polarization curves for mapping out the current transients were chosen as follows: -0.760 V to -0.520 V for the solution with $c_{ion,Fe} = 2.3$ % and -0.700 V to -0.600 V for $c_{ion,Fe} = 27.1$ %.

The potential value at which the transient curve decays the fastest without the appearance of a current more negative than the current value specific for the Cu deposition with constant rate (limiting current) corresponds to the optimum $E_{Cu}^{EC}$ value. As can be inferred from Figs. 2a and 2b, its nominal value is -0.620 V in the case of $c_{ion,Fe} = 2.3$% and -0.660 V in the case of $c_{ion,Fe} = 27.1$ % (for $c_{ion,Fe} = 0$%, its value is -585 mV [13,14]). By plotting these three potential values as a function of the ion concentration of the solution (Fig. 3), the $E_{Cu}^{EC}$ potential values can be determined for intermediate concentrations by interpolation.

According to the above results, contrary to the Ni-Co system, it was found for the Fe-Co system that the value of $E_{Cu}^{EC}$ depends on the relative concentration of $Fe^{2+}$ ions in the solution. This comes from the circumstance that Fe starts to dissolve from the cathode surface at a more negative potential than Co. Because of this difference, if both Fe and Co metals are simultaneously present on the surface with different ratios, the potential value at which the given alloy is neither deposited nor dissolved back into the electrolyte will also depend on the ratio of Fe and Co in the alloy. A further technical problem arises because only the total current flowing through the surface can be measured. Even if the total current is zero, it is possible that Fe is dissolved selectively and, in parallel, Co is deposited at the same rate. However, both the dissolution and the deposition are slow enough for the magnetic layer to be covered with Cu without significant change in the magnetic layer thickness.

*Dependence of magnetic layer composition on the $Fe^{2+}$ ion concentration of the electrolyte.* — By using the $E_{Cu}^{EC}$ potential values determined from the current transients, a sample series was prepared with the help of which the dependence of the Fe concentration $z_{Fe} = c_{Fe} / (c_{Fe} + c_{Co})$ in the magnetic layer of the multilayers on the relative $Fe^{2+}$ ion concentration in the electrolyte could be determined. In this case, the total thickness of each multilayer was 100 nm and the individual layer thicknesses were $d_{Cu} = 3$ nm and $d_{FeCo} = 2.5$ nm. The current density applied in the G pulse used to deposit the magnetic layer was -31.25 mA/cm$^2$. In the above definition of the Fe concentration of the magnetic layer, we have omitted the small amount of Cu unavoidably incorporating during the G pulse. By making the plausible assumption that the Cu current density remains the same during the very short G pulse as it is in the P pulse, the Cu concentration $c_{Cu} = c_{Cu} / (c_{Cu} + c_{Fe} + c_{Co})$ can be estimated by the ratio of the Cu limiting current density (according to Fig. 1, this is about 0.5 mA/cm$^2$ for all Fe concentrations in the baths used for multilayer deposition) and the current density applied in the G pulse as specified above. This yields about 1.5 at.% for the Cu concentration in the magnetic layers.

By measuring the overall multilayer composition in SEM, the Fe concentration $z_{Fe}$ with respect to Co in the magnetic layers of the Fe-Co/Cu multilayers could be determined and these results are shown in Fig. 4.

By taking the nominal layer thicknesses of the Fe-Co/Cu multilayers and accounting also for the 20 nm Cu evaporated underlayer, we can estimate the overall Cu content of the "multilayer + Cu underlayer" system. This way we obtain 62.2 at.% Cu by neglecting the Cu contamination in the magnetic layers and 62.8 at.% Cu for the case when we account for 1.5 at.% Cu in the magnetic layers. The overall Cu content of the multilayers shown in Fig 4 when measured on their substrate in the SEM (thus, including the Cu underlayer as well due to the penetration depth of the EPMA method) was about $61.5 \pm 1.5$ at.%. Both above estimated expected values are well within the experimental error of the composition analysis. These data indicate that the small Cu content in the magnetic layer estimated from the current density ratios in the P and G pulse is in conformity with the measured overall composition. Furthermore, the analysis results can also be interpreted in a manner that the actual layer thicknesses are also fairly close to the nominal values.

The relative $Fe^{2+}$ ion concentration in the electrolyte was increased only up to 40 % (which resulted in 57 at.% Fe in the magnetic layer). At higher Fe molar fractions, a change in the sample surface could already be observed: the surface became matt and spotted. After a few weeks, the oxidation of the surface was visible even by naked eye. This effect was more intense at higher Fe concentrations and this observation is in accordance with the fact that the tendency for corrosion of Fe is the highest among the components of the multilayers studied.

**Structural studies of the ED Fe-Co/Cu multilayers by XRD**

XRD measurements were carried out between $2\Theta$ values of 20° and 120° on selected $Fe_5Co_{95}$(1.5 nm)/Cu($d_{Cu}$) multilayer samples which were prepared from a bath with $c_{ion,Fe} = 5$ % and had Cu layer thicknesses ranging from 0.8 nm to 10.0 nm. The XRD patterns shown in Fig. 5 for the two major reflections (111) and (200) reveal an fcc structure. The peak at 56.3° comes from the Si substrate, with which the individual diffractograms could be precisely positioned with respect to each other. The slight peak immediately below 55° is unidentified but it is certainly a spurious effect only and will be neglected since it is very narrow and has a fixed position so it can hardly originate from the multilayer structure.

It should be noted that the small shoulder on the low-angle side of the (111) peak for the $Fe_5Co_{95}$(1.5 nm)/Cu(0.8 nm) multilayer is not a satellite reflection but it is rather probably due to a small amount of hcp-Co(Fe) phase since a hcp-Co(100) reflection exists at 41.68° [21] whereas the corresponding satellite would be expected to occur at 40.1° at which position no peak can be seen. A small amount of hcp-Co phase has often been observed in ED Co/Cu multilayers [21,24] at Cu layer thicknesses around 1 nm or below and bulk Co-Fe alloys with about 5 at.% Fe can still exhibit the hcp phase. The occurrence of an hcp fraction in an otherwise predominantly fcc multilayer can be explained [24] by the fact that very thin Cu layers are still discontinuous and in the next G pulse, at some uncovered locations the magnetic layer continues to grow on the previously deposited magnetic layer and, thus, a much thicker magnetic layer section can form than the average layer thickness. Without the fcc constraint of a covering Cu layer these Co(Fe) regions will then eventually prefer an hcp structure. As will be shown later, this also shows up in the observed AMR behavior which is a bulk-like characteristic. Although the main Bragg peaks of the hcp-Co(100) and fcc-Co(111) reflections are at practically the same positions [21], the absence of a visible peak at the expected position of the hcp-Co(101) reflection suggests that the main peak even for this sample is probably due to an fcc structure. Due to the discontinuous nature of the very thin Cu layers, the structural coherence along the multilayer thickness gets lost which explains the absence of satellite reflection for the $Fe_5Co_{95}$(1.5 nm)/Cu(0.8 nm) multilayer.

For most of the other multilayers investigated, satellite peaks can be observed on the low-angle side of the main (111) reflection. These satellites are the result of the multilayered structure of the sample and, therefore, are also called superlattice reflections. For a perfectly layered structure, these satellite peaks should show up symmetrically on both sides of the main reflection [25]. The satellite reflections on the higher-angle side of the main peak always have a lower intensity and this might be the reason that in our case they actually fade in the background noise. From the positions of the satellites, the repeat period ($\Lambda$) of a multilayer can be calculated [8,25]. Though mostly having very low intensity, the low-angle side satellites could still be fitted simultaneously with the main peak. By using the fact that the distance of the satellite peaks (for the same order) are equal on both sides [25], from the difference of the sine of their positions the repeat period was calculated. The repeat periods determined in this manner are summarized in Table 1.

According to the last column of Table 1, the obtained experimental $\Lambda$ values are by about 10 to 20 % larger than the preset nominal values, in agreement with our earlier observations on other ED multilayers systems [21,22,24,26].

Figure 6 shows the variation of the positions of the individual peaks as a function of the thickness of the non-magnetic layer. As the Cu layer thickness increases, the position of the main peak moves into the direction of lower angles due to the increase of the average multilayer lattice parameter. In parallel, the satellite peak gets closer and closer to the main peak as the bilayer period ($\Lambda$) increases with increasing Cu layer thickness.

It can also be inferred from Fig. 6 that the broadening (FWHM) of the main XRD peak decreases with increasing Cu layer thickness. The line broadening basically derives from grain size and microstrain effects. Having a constant magnetic layer thickness of 1.5 nm for these multilayers, the number of interfaces per unit multilayer thickness is reduced with increasing Cu layer thickness what results in the reduction of the overall interfacial stresses arising due to the slight lattice mismatch between the magnetic and non-magnetic layers. Thus, the variation of XRD line broadening in these multilayers can probably be assigned to a reduction of the strain contribution due to the relaxed thick Cu layers.

## GMR in ED Fe-Co/Cu multilayers

For the majority of the GMR studies, multilayers with two different $Fe^{2+}$ ion concentrations only were prepared, namely with $c_{ion,Fe} = 2.3$ % and 27.1 %, which resulted in $z_{Fe} = 5$ at. % and 44 at. % Fe in the magnetic layer, respectively. The thickness of the magnetic layers was set to 1.0, 1.5 and 2.0 nm with a proper choice of the length of the G pulse. For all three $d_{FeCo}$ values, several multilayers were prepared with different $d_{Cu}$ values, namely 0.8, 2.4, 4.0, 6.0, 8.0, 10.0 and 12.0 nm. This means 21 multilayers for each of the two selected Fe concentrations whereby the total multilayer thickness was in each case 150 nm.

*Magnetoresistance vs. field curves: separation of the FM and SPM contributions to the GMR.* — Figure 7 shows representative $MR(H)$ curves for $Fe_5Co_{95}$/Cu multilayers with identical magnetic layer thicknesses ($d_{FeCo} = 1.5$ nm) but with two different Cu layer thicknesses ($d_{Cu} = 0.8$ nm and 2.4 nm). These samples belong to the series for which an XRD study was also carried out (Figs. 5 and 6). The upper pair of $MR(H)$ curves in Fig. 7 is for the $Fe_5Co_{95}$(1.5 nm)/Cu(0.8 nm) multilayer: (a) experimental curves and (b) the separated FM and SPM contributions to the measured $MR(H)$ curves, obtained according to the procedure described at the end of the Experimental section.

Since the experimental $MR(H)$ values in Fig. 7a for sufficiently high fields are negative for both the longitudinal and the transverse components, one might think that the sample shows a clear GMR effect due to the layered structure as a consequence of spin-dependent scattering events for electrons travelling between adjacent magnetic layers. However, a closer inspection of Fig. 7b where the separated FM and SPM contributions to the measured magnetoresistance are displayed tells us that the magnetoresistance contribution from spin-dependent scattering events in which FM regions are only involved yields actually a dominant AMR effect ($LMR_{FM} > 0$ and $TMR_{FM} < 0$). The bulk-like dominating AMR effect in the nominally layered structure arises due to the numerous pinholes in the very thin (0.8 nm) spacer layer which is certainly discontinuous providing direct ferromagnetic coupling between adjacent magnetic layers (a hint for this was obtained also from XRD as described in the previous section). It can be seen in Fig. 7b that the GMR effect ($LMR_{SPM} < 0$ and $TMR_{SPM} < 0$) arises mainly due to spin-dependent scattering events in which both SPM and FM regions are involved (scattering events for electrons travelling between two FM regions can be excluded since a $MR(H)$ component proportional to the square of the Langevin function [23] could not be identified).

The results presented in Figs. 7a and 7b clearly underline that, in some cases, how important is to carry out a separation of the FM and SPM contributions to the observed magnetoresistance for establishing the real origin of the obtained MR results. In this particular case, the analysis has revealed that the observed GMR effect does not arise from scattering events between FM regions but between FM and SPM regions only. According to Fig. 7b, the $GMR_{SPM}$ contribution is comparable to the transverse component of the bulk AMR. The formation of a significant amount of SPM regions in ED multilayers often occurs if both kinds of constituent layers are fairly thin [27,28].

According to Figs. 7c and 7d, for a sufficiently thick (here: 2.4 nm) spacer layer, the observed magnetoresistance indeed mainly arises due to spin-dependent scattering events between well separated FM layers although a small SPM contribution still persists. This behavior was characteristic for all multilayers with higher spacer thicknesses just the magnitude of the observed GMR varied with spacer layer thickness as will be shown later.

The $MR(H)$ curves of the ED $Fe_{44}Co_{56}$/Cu multilayers exhibited the same characteristics in that for multilayers with $d_{Cu}$ = 0.8 nm, a predominantly AMR behavior was observed whereas for higher spacer thicknesses the $GMR_{FM}$ was the major contribution with a small additional $GMR_{SPM}$ term. The difference with respect to the $Fe_5Co_{95}$/Cu multilayers was that the observed GMR was much smaller (for very large spacer layer thicknesses, the GMR effect even disappeared) as it will turn out from the data presented below.

*Dependence of GMR on sublayer thicknesses.* — As discussed in the Introduction, in previous studies of the GMR in ED Fe-Co/Cu multilayers [9,10], the Cu deposition potential has not yet been optimized and, thus, the dependence of GMR on true Cu layer thickness could not be established. By measuring the $MR(H)$ curves for all the multilayers listed at the beginning of this section and plotting only the $TMR_{FM}$ component of the GMR, the results presented in Fig. 8 were obtained.

The values of the data points marked with full black circles (●) in Fig. 8 denote samples with $TMR_{FM} = 0$, indicating that these samples show AMR only and no $GMR_{FM}$ contribution (the $GMR_{SPM}$ contribution has no relevance in analyzing the spacer layer thickness dependence of the true multilayer GMR effect). The white-shaded surface regions only indicate thickness parameter ranges with the absence of a GMR effect.

The $TMR_{FM}$ values presented in Fig. 8 indicate that the thickness of the magnetic layer has no significant effect on the multilayer-related GMR in the thickness range explored.

On the contrary, there is a significant variation in the GMR with Cu layer thickness. Similarly to the case of ED Co/Cu [28-30], Ni/Cu [31] and Ni-Co/Cu [32-35] multilayers, no oscillatory GMR could be observed for ED Fe-Co/Cu multilayers either, only a monotonous increase up to a certain $d_{Cu}$ value and a decrease for thicker non-magnetic layers, for the higher Fe content multilayers even vanishing here.

*Dependence of GMR on Fe concentration.* — When comparing Fig. 8a and Fig. 8b, a reduced GMR for higher Fe content in the magnetic layer can be inferred. The dependence of GMR on the Fe content in the magnetic layer was investigated in more detail a further set of 100 nm thick multilayers with $d_{Cu} = 3$ nm and $d_{FeCo} = 2.5$ nm. Figure 9 shows that the GMR decreases monotonically with increasing Fe content, almost vanishing for this particular series at the highest Fe contents explored.

**Comparison of GMR behavior with previously investigated ED Fe-Co/Cu multilayers**

The ED Fe-Co/Cu multilayers for which GMR studies have been reported [9,10] were prepared under quite different conditions than the present ones. The major difference is that in the previous studies the Cu deposition potential was not optimized. The different preparation conditions will not enable us to make a straightforward comparison between the previous and present GMR results; however, to provide a basis for the discussion of the differences in the observed GMR behavior, it may be useful to summarize first briefly the preparation conditions and structural features reported in the previous studies [9,10].

Kakuno et al. [9] investigated the GMR of ED Fe-Co-Cu/Cu multilayers prepared with typically 20 bilayers with a potentiostatic/potentiostatic (P/P) pulse combination from a solution containing 0.9 mol/ℓ $Co^{2+}$ ($CoSO_4$), 0.1 mol/ℓ $Fe^{2+}$ ($FeSO_4(NH_2)SO_4$), 0.005 mol/ℓ $Cu^{2+}$ ($CuSO_4$) and 0.3 mol/ℓ $H_3BO_3$. As substrate, Si(111) wafers covered with an evaporated Cu(111) layer were used. The deposition potentials used were $E_{Fe-Co}(SCE) = -1.24$ V and $E_{Cu}(SCE) = -0.39$ V. No justification for the choice of these potentials was given, it was only noticed that at these electrode potentials the Fe-Co electrodeposits exhibited a very shiny surface whereas the Cu electrodeposits a slightly opaque surface. As to the Fe-Co layer deposition potential, its actual value does not play an important role in determining the properties of the magnetic layer. On the other hand, with reference to our optimization results, the chosen $E_{Cu}(SCE)$ value seems to be by about 0.25 V more positive than the optimum (the similarities of the bath components and their concentrations enables us to assume that the optimum $E_{Cu}$ value would be the same also for their bath as for our one for the particular Fe/Co ionic ratio applied). This means that a strong dissolution of the magnetic layer took place during the Cu deposition pulse and, thus, for each multilayer, the actual magnetic layer thicknesses were less than the nominal preset value and the reverse is true for the Cu layer. According to our previous experience with ED Co/Cu multilayers [13,26], at the Cu deposition potential applied by Kakuno et al. [9], the layer thickness changes due to the magnetic layer dissolution during the Cu deposition pulse may be around 1 nm. This is especially important since the major objective of Kakuno et al. [9] was to study the dependence of GMR on the Cu layer thickness. As to the magnetic layer composition, from a chemical analysis of the multilayers, Kakuno et al. [9] have established that the Fe:Co ratio was 17:83 and the magnetic layer also contained Cu up to 3 at.%. On the basis of an XRD study, it was reported that the multilayers exhibited an fcc(111) texture without evidence of satellite reflections from multilayer periodicity. The (111) diffraction peaks were very broad, indicating small grain sizes. The nearly continuous diffraction rings in the transmission electron microscopy selected area diffraction patterns presented for a multilayer with nominal thicknesses $Fe_{17}Co_{83}$(4 nm)/Cu(4 nm) revealed also a polycrystalline fine-grained character with very low texture. As to the crystal structure, the observed diffraction rings indicated the presence of fcc-Cu, fcc-Co and hcp-Co phases.

In a more recent paper, Tekgül et al. [10] investigated the influence of Fe-content on the GMR of ED Fe-Co/Cu multilayers. Their electrolyte composition was 0.75 M $CoSO_4$, 0.05 M $CuSO_4$, 0.25 M $H_3BO_3$, 0.01 M sulfamic acid and the $Fe^{2+}$ ion concentration was changed from 0 to 0.2 M. The solution pH was 2.5. A P/P pulse combination was used to deposit Fe-Co/Cu multilayers with a total thickness of about 3 μm on a polycrystalline Ti substrate from which the deposits could be mechanically peeled off. The preset nominal Co-Fe and Cu layer thicknesses were kept constant at 6 nm and 4.5 nm, respectively. The applied deposition potentials were $E_{Fe-Co}$(SCE) = -1.5 V and $E_{Cu}$(SCE) = -0.3 V. The Cu deposition potential is even more positive than the one used by Kakuno et al. [9]; therefore, an even stronger dissolution of the magnetic layer and, thus, even larger layer thickness changes with respect to the nominal values can be expected for the multilayers investigated by Tekgül et al. [10]. These authors have also analyzed their multilayers and the resulting Fe/Co ratios in their magnetic layers with varying Fe concentrations in the bath exhibit a fairly good agreement with our data as shown in Fig. 4 (although the comparison may not be completely valid since the current density used in their work for magnetic layer deposition is not known). From the overall multilayer analysis results, Tekgül et al. [10] have also attempted to estimate the Cu content in the magnetic layer which was found to be around 30 at.% for low Fe-content whereas it decreased continuously to about 13 at.% at the highest Fe-content. In this estimate, it was assumed that the actual layer thicknesses are equal to the nominal one, which is definitely not valid for the conditions they applied during multilayer deposition (too positive Cu deposition potential). Thus, due to the larger actual thickness of the Cu layers with respect to the nominal value, the larger measured overall Cu-content was erroneously assigned to the magnetic layer. Therefore, the large Cu-content estimated for the magnetic layer is probably not valid, although a variation of the Cu-content for various Fe-contents cannot be definitely excluded. As to the multilayer structure, XRD patterns revealed an fcc structure with a dominant (111) texture. No satellite reflections were observed.

Our GMR results presented above strongly question previous findings reported by Kakuno et al. [9] on the oscillatory spacer layer thickness dependence of GMR in ED Fe-Co/Cu multilayers. In that work, the Cu deposition potential used was much more positive (namely by 0.25 V) than the optimal value. This must have caused dissolution of the magnetic material during the Cu deposition pulse. Therefore, the thickness of the magnetic layer became thinner while the Cu layer became thicker. As a consequence, as discussed also above, the actual Cu layer thicknesses in Fig. 2 of Ref. 9 are definitely larger (eventually by as much as 1 nm or more) than the values displayed.

As to the magnetoresistance itself, the TMR component was only reported by Kakuno et al. [9]. Due to the magnetic softness of the Fe-Co alloys, the $MR(H)$ curves of the ED Fe-Co/Cu multilayers saturated in magnetic fields around 0.5 kOe for sufficiently thick (2 to 4 nm) magnetic layers. However, for magnetic layer thicknesses at and below 1 nm, the $MR(H)$ curve shape indicated an SPM-type behavior. The high relative magnetic remanence presented hinted at a predominantly FM coupling between adjacent magnetic layers, even for Cu layer thicknesses as high as 4 nm where the magnetoresistance is the highest. This raises the question whether the observed peaks in the Cu layer thickness dependence of the GMR for these multilayers can have any significance. At low Cu layer thicknesses (below about 3 nm), the measured magnetoresistance is anyway so small (0.5 to 1.5 %) that it may arise even from an AMR effect (especially since the TMR component was only reported). Due to the low structural quality of these multilayers (small grain size and lack of XRD satellite peaks), the most that can be established from the reported spacer layer thickness dependence of GMR in Ref. 9 is that the GMR, on the average, increases with Cu layer thickness, just as shown above for our ED Fe-Co/Cu multilayers (see Fig. 6).

In the paper published by Tekgül et al. [10], the potential used was for Cu deposition also more positive than the electrochemically optimal (namely by about 0.3 V). Furthermore, the Cu deposition potential was held constant in the whole Fe concentration range which certainly caused different dissolution rates for every Fe concentration value. The $MR(H)$ curve presented also shows a high saturation field, which is an indication for the presence of a significant amount of SPM regions. Furthermore, no separation of the FM and SPM contributions was carried out, only the $MR(12\ kOe)$ values were reported.

The observed variation of GMR with Fe content of the magnetic layer in our multilayers is in agreement with the results of Tekgül et al. [10] on ED Fe-Co/Cu multilayers. However, the decrease reported by Tekgül is much slower than that was observed in the present study. The reported MR values are also higher. This can be due to the dissolution of the magnetic layer resulting by the Cu deposition potential much more positive than the optimal value (namely by 0.3 V). This could lead to the formation of a large amount SPM regions which dominate the observed GMR. This could be clearly evident by decomposing the measured $MR(H)$ curve in a manner as performed in Fig. 7 above.

**Summary**

In the present work, the structure and magnetoresistance properties were investigated for ED Fe-Co/Cu multilayers. For the first time for this multilayer system, the electrochemically optimal Cu deposition potential was determined at which neither the dissolution of the previously deposited magnetic material nor the codeposition of magnetic atoms with Cu atoms occur. This potential value was found to show a variation with the $Fe^{2+} / [Fe^{2+} + Co^{2+}]$ ionic ratio in the electrolyte. It was found that with increasing Fe content of the magnetic layer, the magnetoresistance decreases, in agreement with the results reported by Tekgül et al. [10] on the same ED multilayer system.

Two multilayer series with 100 nm total thicknesses and with 1.0, 1.5 and 2.0 nm magnetic layer thicknesses were prepared for various Cu layer thicknesses from 0.8 nm to 10.0 nm by electrodeposition on Si wafers with evaporated Cr and Cu underlayer. The composition of the magnetic layers was fixed at $Fe_5Co_{95}$/Cu for one series and at $Fe_{44}Co_{56}$/Cu for the other series.

An XRD study was carried out for the $Fe_5Co_{95}$/Cu series and from the low-angle side satellite peaks originating from superlattice reflections, the bilayer lengths were calculated to be about 10 to 20 % larger than the nominal values as usually observed for ED multilayers [21,22,24,26].

The largest GMR in our ED multilayers was about 5 % and this is well below that reported for physically deposited Fe-Co/Cu multilayers [2-7].

The *GMR* was found to show a maximum at $d_{Cu}$ = 5 nm when the magnetic layer thickness was held constant. By properly decomposing the *GMR* into FM and SPM contributions, it could be concluded that the $GMR_{FM}$ contribution does not exhibit an oscillatory GMR in ED Fe-Co/Cu multilayers as a function of the spacer layer thickness, in agreement with previous observations on various ED multilayers [28-35]. By a detailed analyis of the results of Kakuno et al. [9] about an oscillatory GMR, it could be pointed out that their findings cannot be considered as reliably demonstrating such a behavior.

The thickness of the magnetic layer was found to have no significant effect on the GMR magnitude in the thickness range explored.

**Acknowledgements**

This work was supported by the Hungarian Scientific Research Fund through grant OTKA K 75008. The authors also acknowledge G. Molnár (Institute for Technical Physics and Materials Science, Research Centre for Natural Sciences, HAS) for preparing the evaporated underlayers on the Si substrates.

**Table 1** Summary of repeat period data as determined from the XRD satellites around the main fcc (111) peak

| $\Lambda_{nom}$ (nm) | main peak $2\Theta$ (deg) | satellite $2\Theta$ (deg) | $\Lambda_{exp}$ (nm) | $\dfrac{\Lambda_{exp}}{\Lambda_{nom}}$ |
|---|---|---|---|---|
| 3.9 | 43.49 | 41.64 | 5.11 | 1.31 |
| 5.5 | 43.47 | 42.14 | 7.12 | 1.29 |
| 7.5 | 43.41 | 42.39 | 9.30 | 1.24 |
| 9.5 | 43.36 | 42.46 | 10.6 | 1.12 |
| 11.5 | 43.38 | 42.62 | 12.4 | 1.08 |

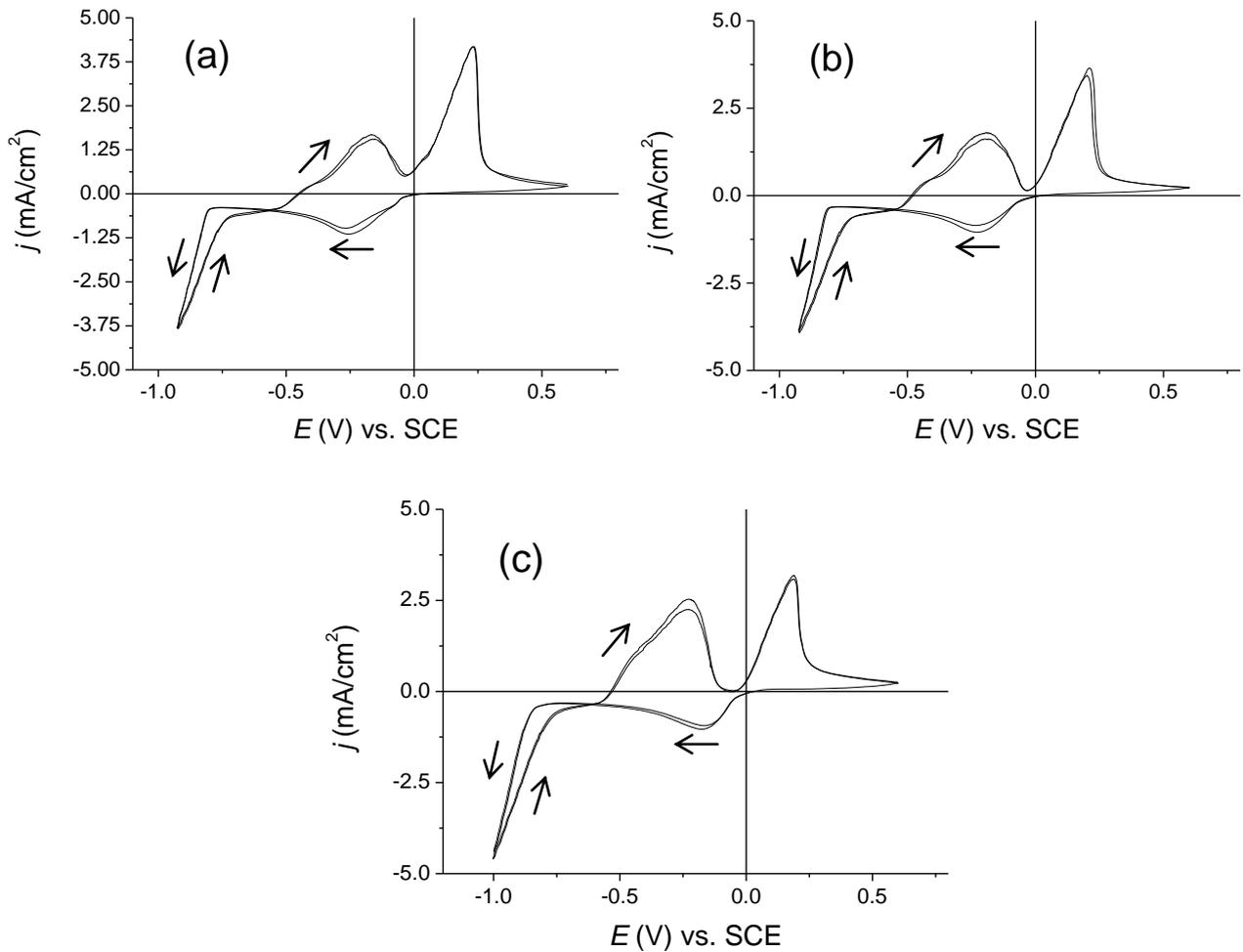

Fig. 1 Cyclic voltammograms for the Co-Cu solution and for two Fe-Co-Cu solutions with different $Fe^{2+}$-concentrations ($c_{ion,Fe}$): (a) $c_{ion,Fe}$ = 0 %; (b) $c_{ion,Fe}$ = 2.3 %; (c) $c_{ion,Fe}$ = 27.1 %.

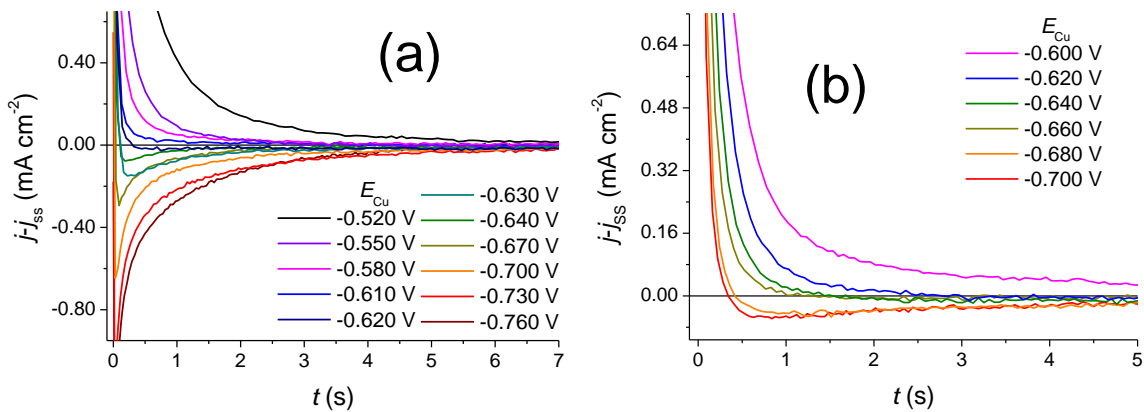

Fig. 2 Current transients for the two Fe-Co-Cu electrolytes with two different $Fe^{2+}$ ion concentrations $c_{ion,Fe}$ = 2.3 % (a) and $c_{ion,Fe}$ = 27.1 % (b) in the selected potential range to find the optimal Cu deposition potential $E_{Cu}^{EC}$. The potential range was determined from the cyclic voltammograms shown in Figs. 1(a) and 1(b). The transients were measured in the 20th cycle which can be considered as steady-state. $j$-$j_{ss}$: the difference between the measured and the steady-state current densities, t: time elapsed from the beginning of the pulse, $d_{FeCo}$ = 2.5 nm, $d_{Cu}$ = 3.0 nm.

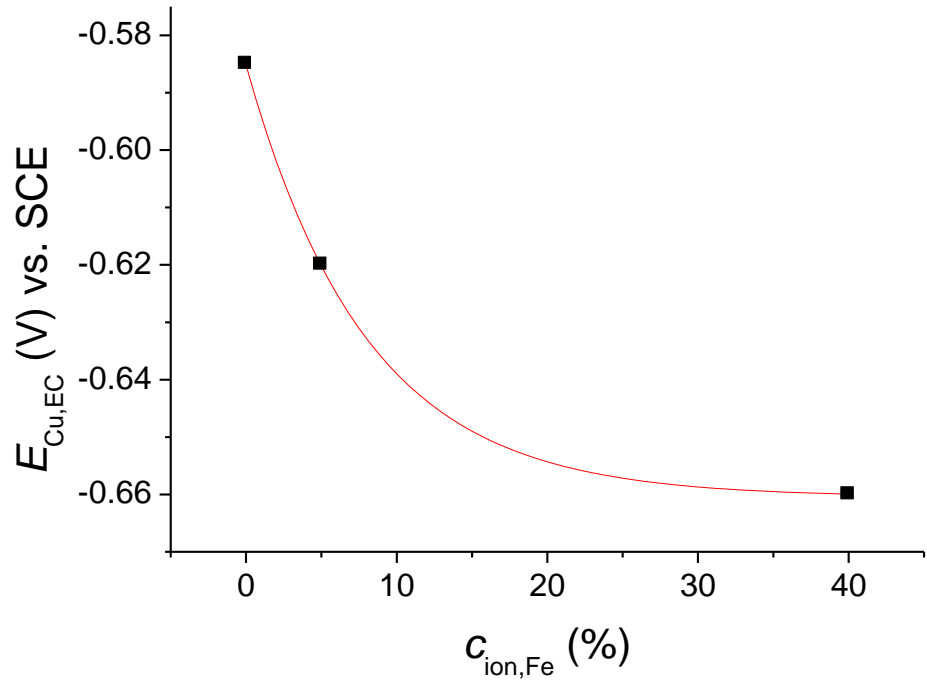

Fig. 3  Optimized Cu deposition potential (■) as a function of the $Fe^{2+}$ concentration in the solution as determined from the current transients shown in Fig. 2. The continuous line is an interpolation for intermediate concentrations.

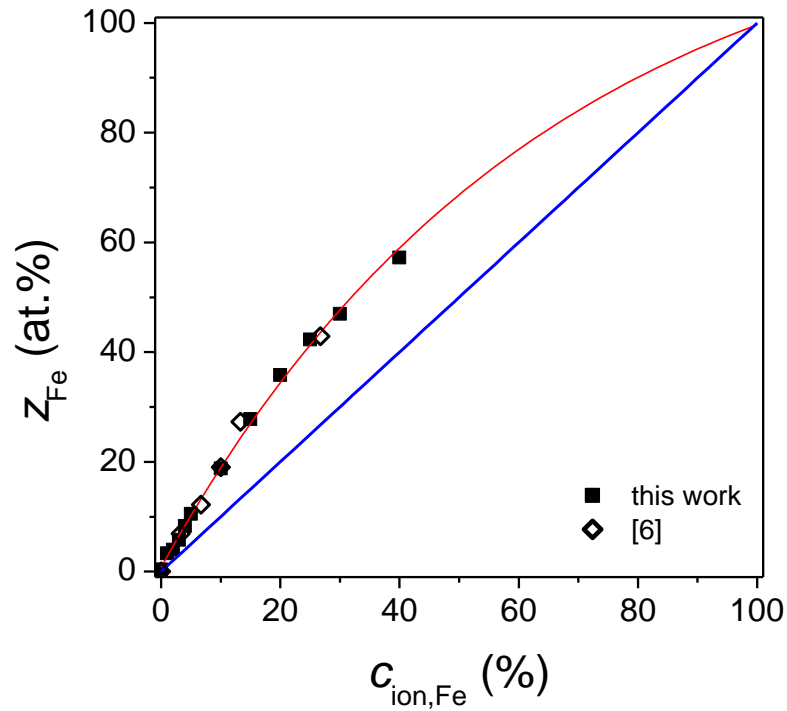

Fig. 4  Dependence of the Fe concentration $z_{Fe}$ (■) of the magnetic layers in the ED Fe-Co/Cu multilayers on the relative $Fe^{2+}$ ion concentration $c_{ion,Fe}$ of the electrolyte. The thick blue line with a slope equal to 1 corresponds to the conditions of equilibrium codeposition. The thin red line represents the Fe concentrations of the magnetic layer extrapolated to higher $Fe^{2+}$ ion concentrations and helps interpolate in the region measured. Symbols ◇ represent data from Ref. 10 (see text for details).

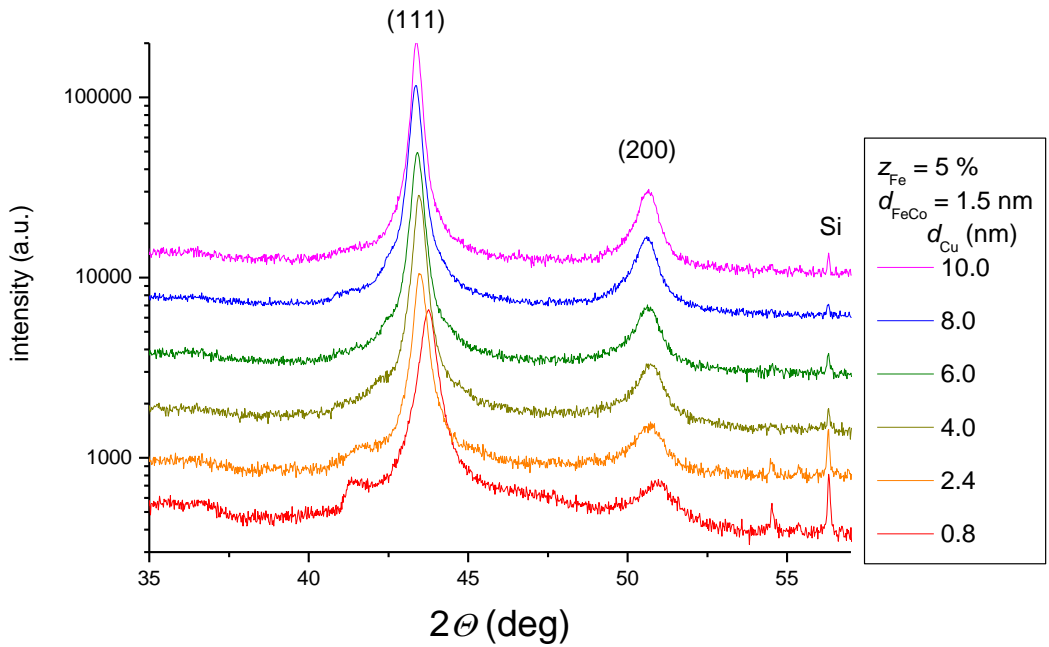

Fig. 5  X-ray diffractograms of Fe$_5$Co$_{95}$/Cu multilayers with different Cu layer thicknesses. The magnetic layer thickness and the total multilayer thickness were fixed at 1.5 nm and 100 nm, respectively. The patterns were shifted along the ordinate axis for clarity.

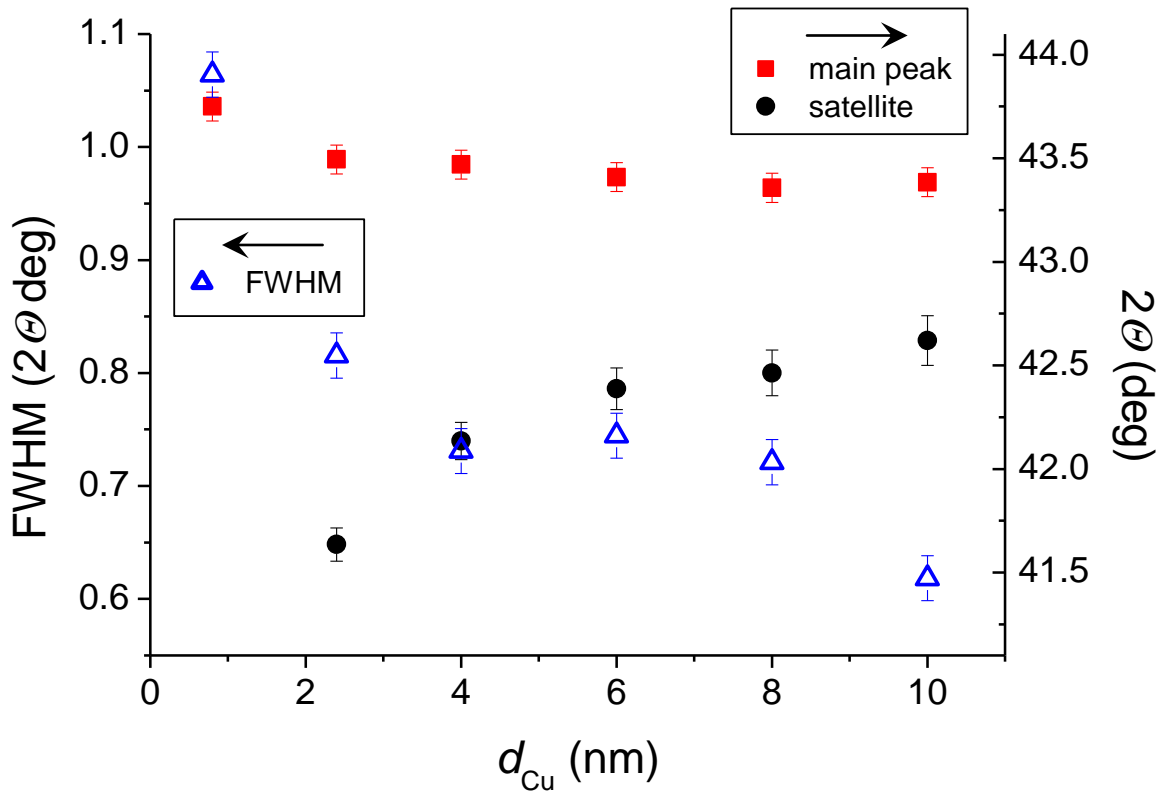

Fig. 6  The $2\Theta$ positions of the (111) peak and its low-angle satellite reflection for Fe$_5$Co$_{95}$/Cu multilayers with different Cu layer thicknesses. The magnetic layer thickness and the total multilayer thickness were fixed at 1.5 nm and 100 nm, respectively. The variation of the width of the main peak as characterized by its FWHM value is also plotted.

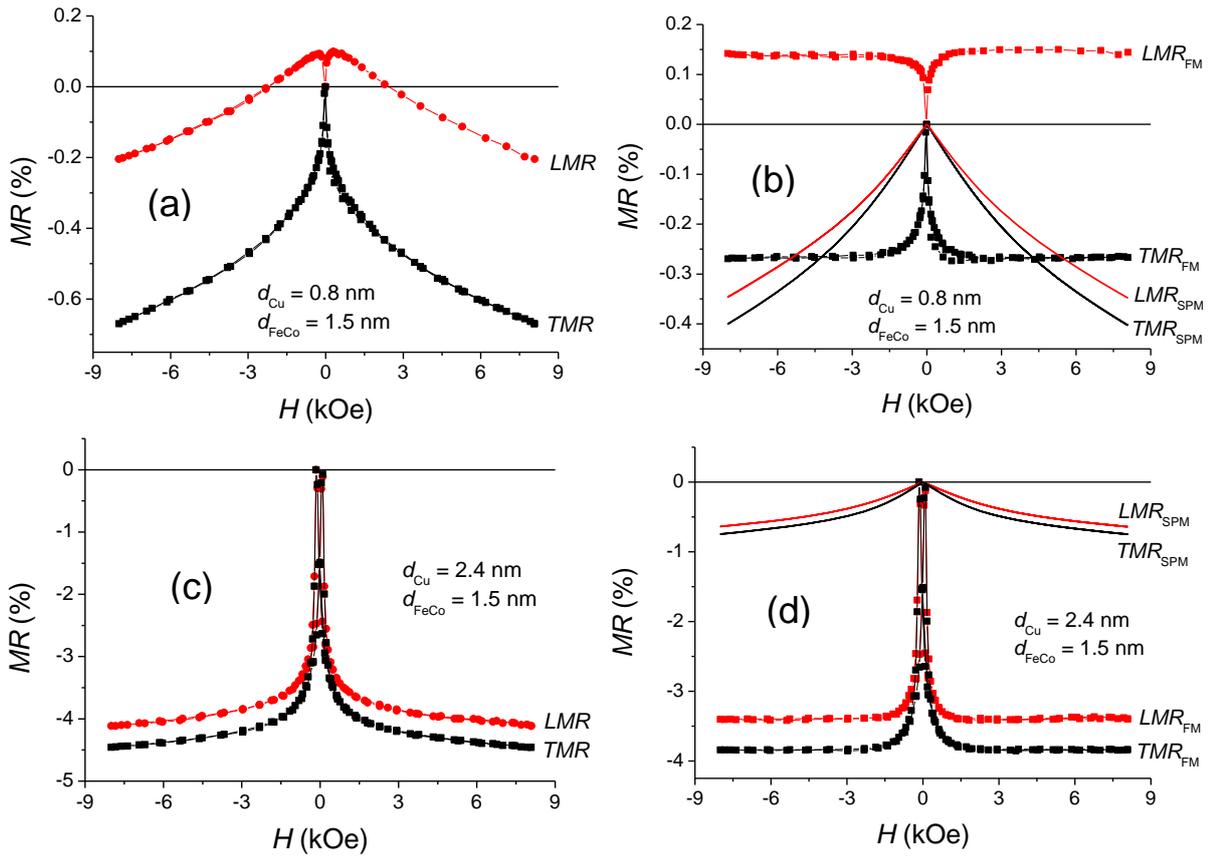

Fig. 7 Longitudinal and transverse *MR(H)* curves for Fe$_5$Co$_{95}$/Cu multilayers with a total multilayer thickness of 150 nm for two different Cu layer thicknesses: (a) and (b): $d_{Cu}$ = 0.8 nm; (c) and (d): $d_{Cu}$ = 2.4 nm; the magnetic layer thickness was 1.5 nm for both samples. For each multilayer, the left graph displays the experimental data whereas the right graph shows the results of decomposing the measured magnetoresistance into FM and SPM contributions.

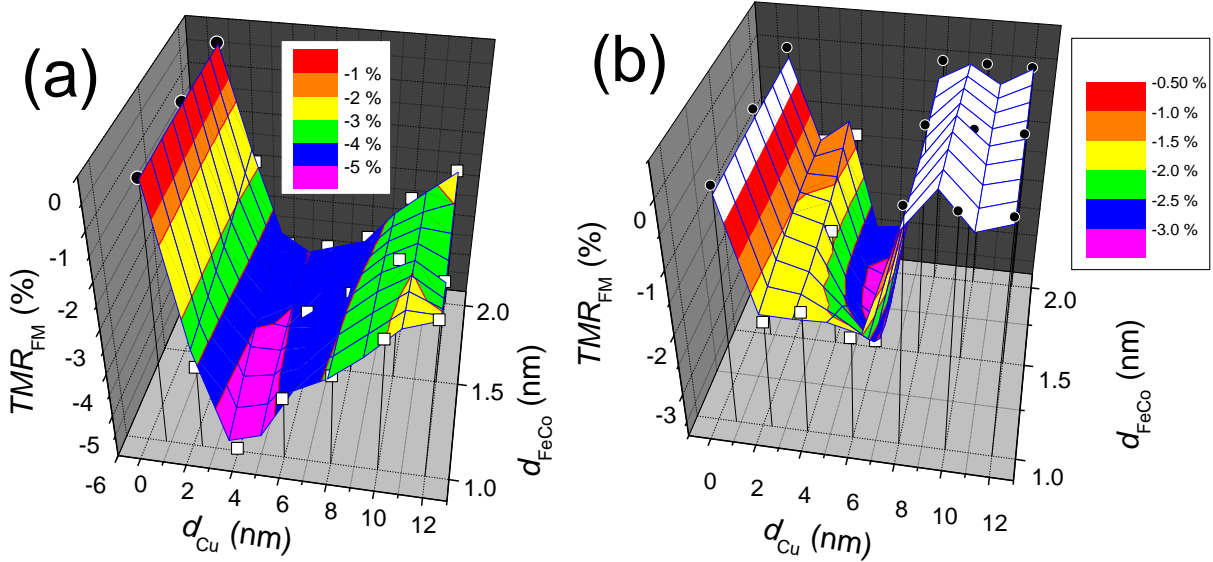

Fig. 8  Dependence of the FM component of the *TMR* (*TMR*$_{FM}$) on the thickness of the magnetic layer ($d_{FeCo}$) and the Cu layer ($d_{Cu}$) in our ED Fe-Co/Cu multilayers. (a) $z_{Fe}$ = 5 % (b) $z_{Fe}$ = 44 %. Empty squares (□) mark the *TMR*$_{FM}$ component of samples showing GMR and black full circles (●) with zero value represent the samples showing AMR. The surfaces over the data points obtained by fitting are only displayed to better show the evolution of the MR data with layer thicknesses. The white-shaded surface regions only indicate thickness parameter ranges with the absence of a GMR effect, regardless the artificial wavy shape due to fitting.

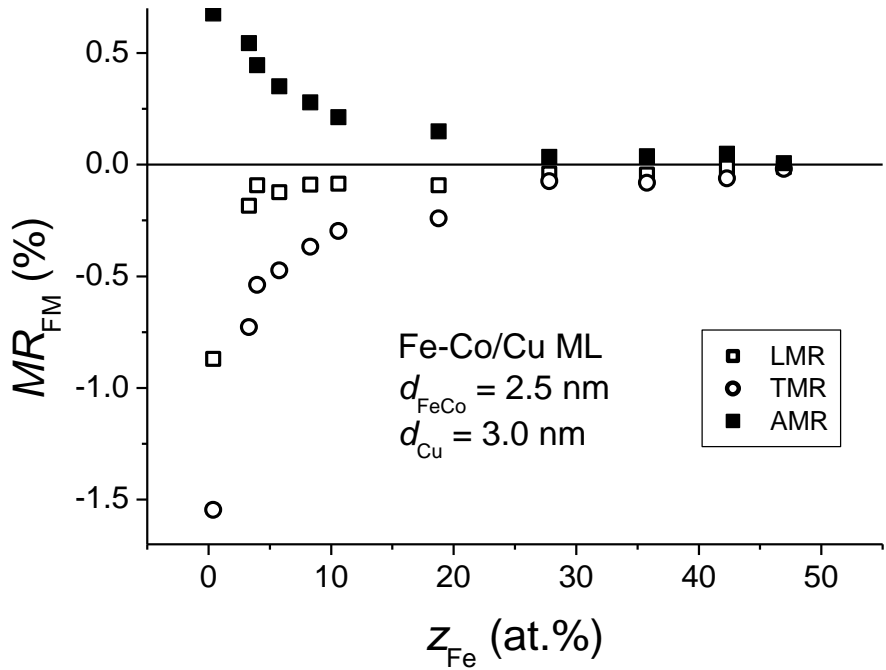

Fig. 9  Dependence of the longitudinal (*LMR*, □) and transverse (*TMR*, ○) component of the *MR* and their difference, the *AMR* (■) on the Fe content of the magnetic layer ($z_{Fe}$). The total thickness (Σ$d$) of the multilayers was 100 nm, the thickness of the magnetic layer ($d_{FeCo}$) and the Cu layer ($d_{Cu}$) was 2.5 and 3.0 nm, respectively.